\documentclass[superscriptaddress,twocolumn,showpacs,amsmath,amssymb]{revtex4}
\usepackage[dvips]{graphicx}

\def\be{\begin{equation}}
\def\ee{\end{equation}}
\def\bea{\begin{eqnarray}}
\def\eea{\end{eqnarray}}
\def\br{{\bf r}}
\def\bp{{\bf p}}
\def\rmC{{\rm C}}

\def\rmpC{{\rm pC}}
\def\rmpp{{\rm pp}}
\def\rmE1{{\rm E1}}
\renewcommand{\vec}[1]{\mbox{\boldmath $#1$}}

\begin{document}

\title{
Effect of proton-proton Coulomb repulsion on soft dipole excitations 
of light proton-rich nuclei}

\author{T. Oishi}
\author{K. Hagino}
\affiliation{
Department of Physics,  Tohoku University,  Sendai,  980-8578,  Japan}

\author{H. Sagawa}
\affiliation{
Center for Mathematical Sciences,  University of Aizu,  Aizu-
Wakamatsu,  Fukushima 965-8560,  Japan}


\begin{abstract}
We perform three-body model calculations for soft dipole 
excitations of the proton-rich Borromean nucleus $^{17}$Ne. 
To this end, we assume that 
$^{17}$Ne takes the 
$^{15}$O+p+p structure, in which the two valence 
protons are excited from the $0^+$ ground state configuration to $1^-$ 
continuum states. 
We employ 
a density-dependent contact force for the nuclear part of the 
pairing interaction, and 
discretize the continuum states with the box boundary condition. 
We show by explicitly including the Coulomb interaction between 
the valence protons that 
the Coulomb repulsion does not significantly alter the E1 strength distribution. 
We point out that the effect of the Coulomb repulsion in fact can be well 
simulated by renormalizing the nuclear pairing interaction. 
\end{abstract}


\maketitle


Properties of unstable nuclei with large excess of proton or neutron 
are one of the most important current topics of nuclear physics. 
In several radioactive beam facilities in the world, 
many unstable nuclei far from the $\beta$-stability line 
have been discovered\cite{T8588,Oza94,Oza01,Jon04}. 
In particular, 
neutron-rich unstable nuclei have been extensively studied 
and some exotic features have been observed. 
This includes a large concentration of the dipole strength 
distribution at low energies\cite{F04,N06}, 
referred to as a ``soft dipole excitation''. 
This type of excitation is naively understood as an oscillation 
between weakly bound valence nucleon(s) 
and  the core 
nucleus\cite{IMKT10,Sag95}. 
The relation between the soft dipole excitation and 
a largely extended density distribution, that is, 
a halo or skin property has been 
discussed for light neutron-rich nuclei, both 
experimentally \cite{F04,N06,K89,A90,Shi95,A99,L01,P03} and 
theoretically \cite{HJ87,BB88,BE91,EB92,EBH97,EHMS07}. 
Recently, the neutron halo structure has been discussed also in 
the $^{31}$Ne nucleus, 
based on the measured large Coulomb break-up cross sections\cite{N09}. 

In contrast to neutron-rich nuclei, 
proton-rich nuclei have been less studied. 
It has not been fully clarified whether similar 
exotic features are present also in proton-rich nuclei. 
For instance, a proton halo structure 
in {\it e.g.,} $^8$B, $^{12}$N, $^{17}$F, and $^{17}$Ne 
has been discussed\cite{Sag95,ZT95,GPZ05,GLSZ06}, 
but no clear evidence has been obtained so far. 

In order to investigate proton-rich unstable nuclei 
and discuss their similarities and differences 
to 
neutron-rich nuclei, 
it is indispensable to assess the 
effect of the Coulomb repulsion between valence protons. 
In the previous work, we analyzed the ground state properties 
of $^{17}$Ne using a three-body model \cite{OHS10}. 
We have shown that 
the effect of the Coulomb repulsion is weak enough and 
the two valence protons in the ground state of $^{17}$Ne 
have a spatially compact configuration, that is, 
the diproton correlation, 
similar to a dineutron correlation in neutron-rich Borromean 
nuclei \cite{HS0507,MMS05,M06,PSS07,KEHSS09}. 
We have also shown that the effect of the Coulomb 
interaction between the valence protons can be well 
accounted for by renormalizing the nuclear interaction. 
A similar conclusion was achieved recently also by Nakada and Yamagami,  
who performed 
Hartree-Fock-Bogoliubov (HFB) calculations for $N$=20,28,50,82, and 
126 isotones \cite{NY11}. 

In this paper, 
as an continuation of the previous study, 
we discuss the effect of the Coulomb repulsion 
on excited states of $^{17}$Ne. 
Our interest is to investigate whether a similar renormalization 
for the Coulomb repulsion works also for 
the soft dipole excitation, 
which plays an important role in 
the astrophysical two-proton capture 
on $^{15}$O\cite{GLSZ06}. 


We assume the $^{17}$Ne nucleus as a three-body system composed of 
an inert spherical core nucleus $^{15}$O and two valence protons. 
The three-body Hamiltonian in the three-body rest frame reads 
\bea
 H &=& h^{(1)}+h^{(2)}+\frac{\bp_1 \cdot \bp_2}{A_{\rmC}m}
      +V_\rmpp(\br_1,\br_2),
\label{eq:3bh} \\
 h^{(i)}
   &=& \frac{\bp_i^2}{2\mu}+V_{\rmpC}(\br_i),
\label{eq:sph}
\eea
where $m$ and $A_{\rmC}$ are the nucleon mass and the mass number of 
the core nucleus, respectively. 
$h^{(i)}$ is the single-particle (s.p.) Hamiltonian for a valence 
proton, in which $\mu=mA_{\rmC}/(A_{\rmC}+1)$ is the reduced mass 
and $V_{\rmpC}$ is the potential between the proton and the 
core nucleus. 
The third term in Eq.(\ref{eq:3bh}) is a 
two-body part of the recoil kinetic energy of the core nucleus. 
For the proton-core potential $V_{\rmpC}$, 
we employ a Woods-Saxon (WS) plus Coulomb potential 
in the same manner as in the previous work \cite{OHS10}. 
We use the same parameters for the WS potential as in those listed 
in Ref.\cite{OHS10}. 

We solve the three-body Hamiltonian, Eq. (\ref{eq:3bh}), 
by expanding the wave function 
on the uncorrelated basis as, 
\be
 \Psi^{(J,M)}(\br_1,\br_2)
 =  \sum_{k_1 \leq k_2} \alpha_{k_1,k_2}
       \tilde{\psi}^{(J,M)}_{k_1,k_2} (\br_1,\br_2),
\label{eq:WF} 
\ee
where
\bea
 \tilde{\psi}^{(J,M)}_{k_1,k_2} (\br_1,\br_2)
 & = & \nonumber \frac{1}{\sqrt{2}}
       \left[ 1-(-)^{j_1+j_2-J}\delta_{k_1,k_2} \right]^{-1} \\
 & \times & \nonumber \sum_{m_1,m_2} \langle j_1 m_1 j_2 m_2\mid J M \rangle \\
 & \times & \nonumber \left[ \right. 
           \phi_{k_1,m_1}(\br_1) \phi_{k_2,m_2}(\br_2) \\
 & & -\phi_{k_2,m_2}(\br_1) \phi_{k_1,m_1}(\br_2) \left. \right].
\label{eq:basis}
\eea
Here, $\phi_{km}(\br)$ is a s.p. wave function with 
$k=(n,l,j)$, while 
$J$ and $M$ are the total angular momentum of 
the two-proton subsystem and its $z$ component, respectively. 
$\alpha_{k_1,k_2}$ is the expansion coefficient. 
The summation in Eq. (\ref{eq:WF}) is restricted to those combinations which 
satisfy $\pi=(-)^{l_1+l_2}$ for a state with parity $\pi$. 

In the actual calculations shown below, 
we include the s.p. angular momentum $l$ up to 5. 
We have confirmed that our results do not change 
significantly even if we include up to a larger value of $l$. 
In order to take into account the effect of the 
Pauli principle, we explicitly exclude the 
$1s_{1/2}$, $1p_{3/2}$, and $1p_{1/2}$ states from Eq. (\ref{eq:WF}), which are 
occupied by the protons in the core nucleus. 

For the pairing interaction $V_\rmpp$, 
we assume a density-dependent contact 
interaction \cite{BE91,EBH97,HS0507} together with 
the Coulomb repulsion, 
$V_\rmpp=V_\rmpp^{(N)}+V_\rmpp^{(C)}$, 
as in the previous work \cite{OHS10}. 
We take a cutoff energy $E_{\rm cut}=30$ MeV and include those configurations 
which satisfy 
\be
 \epsilon_{n_1l_1j_1}+\epsilon_{n_2l_2j_2}\leq
 \frac{A_{\rmC}+1}{A_{\rmC}}\,E_{\rm cut},
\label{eq:Ecut}
\ee
where $\epsilon_{nlj}$ is a s.p. energy \cite{EBH97}. 
Within this truncated space, we determine the strength of  
the nuclear part of the pairing interaction, $V_\rmpp^{(N)}$, 
using the empirical value of the neutron-neutron scattering length, 
$a_{\rm nn}=-18.5$ fm \cite{EBH97}. 
The parameters for the density dependence are adjusted so as to 
reproduce the experimental value of 
the two-proton separation energy of $^{17}$Ne, $S_{\rm 2p}=0.944$ MeV. 

In our calculations, the continuum s.p. spectra are discretized 
within a box of $R_{\rm box}$=30 fm. 
Thus, energies of the two-proton $1^-$ states as well as the E1 
strength distribution are also discretized. 
The E1 strength function from the ground state 
is defined as
\be
 S(E) = \sum_{i} \mu_i \, \delta(E-\hbar\omega_i)
\ee
where $\hbar\omega_i=E_i-E_{\rm g.s.}$, $E_{\rm g.s.}$ being the energy of the 
ground state, 
and $\mu_i$ is the $B(\rmE1)$ strength for 
$i$-th $1^-$ two-proton state, 
\be
 \mu_i
 =3 \left| \left< \Psi^{(1,0)}_i \mid \hat{D}_0 \mid \Psi_{\rm g.s.}
    \right> \right|^2,
\ee
with the E1 operator given by 
\be
 \hat{D}_{\mu}
 = e\left( \frac{A_{\rmC}-Z_{\rmC}}{A_{\rmC}+2} \right)
     \left[ r_1Y_{1\mu}(\hat{\br}_1)+r_2Y_{1\mu}(\hat{\br}_2) \right].  
\ee
Using the strength function $S(E)$, we can also calculate the $k$-th moment of 
energy defined as  
\be
 S_k = \int dE \,E^k S(E) = \sum_i (\hbar\omega_i)^k \,\mu_i.
\ee
Notice that $S_0$ and $S_1$ correspond to the direct and energy-weighted 
sum of $dB(\rmE1)/dE$, respectively. 

From the completeness of the $1^-$ basis, we can estimate the sum-rule-values as 
\bea
 S_{0,{\rm SR}}
 &=& \frac{3}{\pi} e^2 \left( \frac{A_{\rmC}-Z_{\rmC}}{A_{\rmC}+2} \right)^2
     \left< \Psi_{\rm g.s.} | r^2_{\rm 2N-C} | \Psi_{\rm g.s.} \right>, \\
 S_{1,{\rm SR}}
 &=& \frac{9}{4\pi} e^2 \left( \frac{A_{\rmC}-Z_{\rmC}}{A_{\rmC}+2} \right)^2
     \frac{A_{\rmC}+2}{A_{\rmC}m} \hbar^2,
\eea
where $\vec{r}_{\rm 2N-C}=(\br_1+\br_2)/2$ is the distance between 
the center of mass of the two valence protons and 
the core nucleus. 
For the core nucleus $^{15}$O ($A_{\rmC}=15$ and $Z_{\rmC}=8$), 
we obtain $S_{0,{\rm SR}}=1.49$ $e^2$fm$^2$ 
and $S_{1,\rm SR}=5.69$ $e^2$fm$^2$MeV. 
Due to the Pauli forbidden transitions, the actual value of 
$S_0$ is smaller than $S_{0,{\rm SR}}$, while 
$S_1$ is larger than $S_{1,{\rm SR}}$.


\begin{table}[h]
\caption{
The results for the soft dipole excitations of 
$^{17}$Ne obtained with the three-body model of $^{15}$O+p+p.
$S_0$ and $S_1$ are the non-energy weighted sum rule and the energy 
weighted sum rule, respectively. $E_{\rm cent}=S_1/S_0$ is the 
centroid energy of the dipole strength distribution.
$\delta E_{\rm cent}$ is a shift of the centroid energy with respect to the 
result of the exact treatment of the Coulomb interaction. 
}
\begin{center}
\begin{tabular}{c c c c c} 
\hline \hline
pairing & $S_0$ & $S_1$ & $E_{\rm cent}$ 
& $\delta E_{\rm cent}$ \\ 
& ($e^2$fm$^2$) & ($e^2$fm$^2$MeV) & (MeV) 
& (MeV)\\ \hline
Nucl. + Coul. & 1.206 & 11.02 & 9.140 & 0 \\
Nucl. only & 1.206 & 10.45 & 8.666 & $-$0.47 \\
Ren. Nucl. & 1.205 & 10.86 & 9.017 & $-$0.12 \\
No pairing & 1.206 & 15.50 & 12.86 & 3.72 \\
\hline \hline
\end{tabular}
\label{tb:tb1}
\end{center}
\end{table}

\begin{figure}[t]
\begin{center}
\includegraphics[width=8.0cm, height=6.0cm]{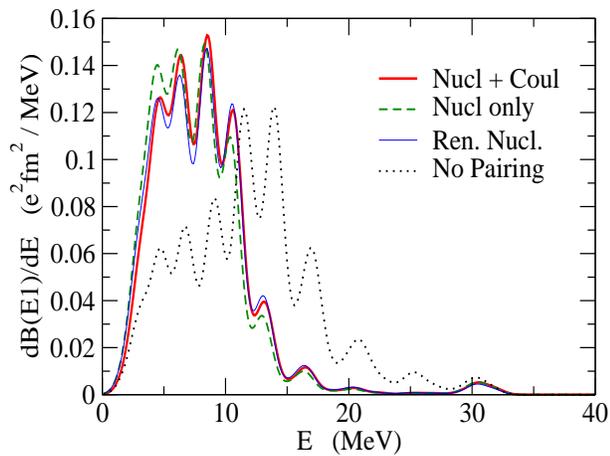}
\caption{(Color online)
Comparison of the 
E1 strength distributions for $^{17}$Ne 
obtained with several treatments for the Coulomb 
interaction between the valence protons. 
The solid line is obtained by fully including the Coulomb interaction, 
while the dashed line is obtained by switching off the Coulomb repulsion. 
The thin solid line is the result of the renormalized nuclear interaction, 
which is readjusted to reproduce the ground state energy without the 
Coulomb interaction. The dotted line denotes the result without 
any pairing interaction. 
These distributions are 
smeared with the Cauchy-Lorentz function, Eq. (\ref{eq:CL}), 
with $\Gamma = 1.0$ MeV. 
}
\label{fig:fig1}
\end{center}
\end{figure}

Our main results are summarized in 
Fig. \ref{fig:fig1} and Table \ref{tb:tb1}. 
For a plotting purpose, 
we smear the 
Dirac delta function in the E1 distribution $S(E)$ with a 
Cauchy-Lorentz function 
\be
 \frac{dB(\rmE1)}{dE}
 = \sum_{i} \mu_i \, \frac{\Gamma}{\pi} \frac{1}{(E-\hbar\omega_i)^2+\Gamma^2} 
 \label{eq:CL}
\ee
with the width parameter $\Gamma$ of $1.0$ MeV. 
In order to discuss the effect of the Coulomb part of 
the pairing interaction, 
we also perform the calculations with two other treatments for 
the Coulomb interaction. 
One is to switch off the Coulomb interaction (``Nucl. only''), 
that is, $V_\rmpp=V_\rmpp^{(N)}$, 
keeping the same values for the parameters of the 
nuclear pairing interaction $V_\rmpp^{(N)}$ as in the full 
calculation (``Nucl.+Coul.''). 
The other is again to use the nuclear interaction only, but renormalize 
the parameters, 
that is, $V_\rmpp=\tilde{V}_\rmpp^{(N)}$ (``Ren. Nucl.''). 
To renormalize the interaction, 
we use the empirical proton-proton scattering length, 
$a_{\rm pp}=-7.81$ fm, 
instead of the neutron-neutron scattering length $a_{\rm nn}$, 
and determine the other parameters so that the two-proton separation 
energy $S_{\rm 2p}$ is reproduced. This leads to about 10.0\% reduction 
of the strength of the pairing interaction. Notice that this value for the 
reduction factor is consistent with 
the finding of Ref. \cite{NY11}. 
See Ref.\cite{OHS10} for further details of the procedure. 
For a comparison, we also show the results without any pairing interaction. 
These treatments for the Coulomb interaction are applied only to the excited 
states, while the same ground state wave function is used for all the cases. 
That is, the ground state is calculated with the full treatment of the Coulomb 
interaction, that yields 
76\% of $(d_{5/2})^2$ and 16\% of $(s_{1/2})^2$. 
This ground state is used also for the no-pairing calculation for the 
dipole excitations. 
(These values for the occupation probabilities are 
slightly different from those in Ref.\cite{OHS10}, as we 
use a smaller $E_{\rm cut}$ in this paper. 
We have confirmed that the dipole strength distribution does not change much 
even though we use the smaller value of $E_{\rm cut}$.) 
Notice that 
our results for $S_0$ are consistent with 
the result of Grigorenko {\it et al} \cite{GLSZ06}, 
that is, $S_0=1.56$ $e^2$fm$^2$ for $(s_{1/2})^2$=48\% 
and $S_0=1.07$ $e^2$fm$^2$ for $(s_{1/2})^2$=5\%. 
Table \ref{tb:tb1} also lists the centroid energy defined 
as $E_{\rm cent} \equiv S_1/S_0$ \cite{SSAR06}, 
and its relative value with respect to the result of the full treatment of the 
Coulomb interaction. 

From Fig. \ref{fig:fig1} and Table \ref{tb:tb1}, 
we can see that the pairing interaction shifts considerably the 
E1 strength distribution towards the low energy region, similarly 
to the dipole distribution in 
neutron-rich nuclei \cite{EHMS07,N06}. 
This large shift of the E1 strength distribution originates mainly from the 
nuclear part of the pairing interaction. If the Coulomb part is switched off, 
the strength distribution is shifted only slightly, as is shown 
in Fig. \ref{fig:fig1} by the dashed line. 
The shift of the centroid energy, $\delta E_{\rm cent}$, is 
only $-0.47$MeV. 
The sign and the magnitude of $\delta E_{\rm cent}$ is consistent with 
the result of Hartree-Fock+RPA calculations for 
medium-heavy nuclei shown in 
Ref. \cite{SSAR06}, although 
$\delta E_{\rm cent}$ for the soft dipole excitation in $^{17}$Ne is 
somewhat larger. 

The result with the renormalized nuclear pairing interaction is shown by 
the thin solid line in Fig. \ref{fig:fig1}. 
As one can see, 
the result of the full calculation (the thick solid line) 
is well reproduced by this prescription. 
We can thus conclude that the renormalization works well 
not only for the ground state \cite{OHS10,NY11}, 
but also for the dipole excitations. 

In summary, 
we discussed the influence of the Coulomb repulsion between 
the valence protons upon the soft dipole excitation in $^{17}$Ne. 
We showed that the effect of the Coulomb repulsion is 
so weak that the main feature of the dipole response is similar 
between neutron-rich and proton-rich weakly bound nuclei. 
We also showed that the effect of the Coulomb interaction can be well 
mocked up by renormalizing the nuclear pairing interaction. 
This renormalization works both for the ground state and for the dipole 
excitations, and thus from a practical point of view one can 
use a renormalized pairing interaction in order to understand the 
structure of proton-rich nuclei. 

One of the current topics of proton-rich nuclei is two-proton 
radioactivity. 
Bertulani, Hussein, and Verde argued that the final 
state interaction plays an important role in discussing the energy and 
the angular correlations in a two-proton emission process \cite{BHV08}. 
It would be an interesting future work to investigate how the 
renormalization works for those correlations. 
A work towards this direction is now in progress. 

\bigskip

This work was supported 
by the Grant-in-Aid for Scientific Research (C), 
Contract No. 22540262 and 20540277 from the Japan Society for 
the Promotion of Science. 

\end{document}